\documentclass[preprint2]{aastex}

\slugcomment{Resubmitted to PASP, September 2007.}

\shorttitle{Period Changes of RT~Aurigae}
\shortauthors{Turner et al.}

\begin{document}

\title{The Period Changes of the Cepheid RT~Aurigae}

\author{David G. Turner}
\affil{Department of Astronomy and Physics, Saint Mary's University,
    Halifax, Nova Scotia B3H 3C3, Canada}
\email{turner@ap.smu.ca}

\author{Ivan S. Bryukhanov, Igor I. Balyuk, Alexey M. Gain, Roman A. 
    Grabovsky, Valery D. Grigorenko, Igor V. Klochko, Attila Kosa-Kiss, 
    Alexey S. Kosinsky, Ivan J. Kushmar, Vyacheslav T. Mamedov, Natalya A. 
    Narkevich, Andrey J. Pogosyants, Andrey S. Semenyuta, Ivan M. Sergey, 
    Vladimir V. Schukin, Jury B. Strigelsky, and Valentina G. Tamello}
\affil{Group ``Betelgeuse,'' Republic Center of Technical Creativity of 
    Pupils, 12 Makaionak Street, Minsk 220023, Belarus}
\email{betelgeize\underline{ }astro@mail.ru}

\author{David J. Lane and Daniel J. Majaess}
\affil{Department of Astronomy and Physics, Saint Mary's University,
    Halifax, Nova Scotia B3H 3C3, Canada}

\begin{abstract}
Observations of the light curve for the 3.7-day Cepheid RT~Aur both before 
and since 1980 indicate that the variable is undergoing an overall period 
increase, amounting to $+0.082 \pm 0.012$ s yr$^{-1}$, rather than a period 
decrease, as implied by all observations prior to 1980. Superposed on the 
star's O--C variations is a sinusoidal trend that cannot be attributed to 
random fluctuations in pulsation period. Rather, it appears to arise from 
light travel time effects in a binary system. The derived orbital period 
for the system is $P = 26,429\pm89$ days ($72.36\pm0.24$ years). The 
inferred orbital parameters from the O--C residuals differ from those 
indicated by existing radial velocity data. The latter imply the most 
reasonable results, namely $a_1{\rm sin}i = 9.09(\pm1.81)\times10^8$ km and 
a minimum secondary mass of $M_2 = 1.15\pm0.25\;M_{\sun}$. Continued 
monitoring of the brightness and radial velocity changes in the Cepheid are 
necessary to confirm the long-term trend and to provide data for a proper 
spectroscopic solution to the orbit. \end{abstract}

\keywords{Stars}

\section{Introduction}

Every well-studied Cepheid undergoes changes in pulsation period: some 
rapidly, others extremely slowly, and $\sim10\%$ in irregular fashion, 
attributable to random fluctuations in pulsation period, generally 
superposed upon parabolic evolutionary trends, e.g. SV~Vul \citep{tb04}. 
For the large majority the effect can be attributed directly to gradual 
changes in mean radius as post main-sequence stars of $3-20\;M_\sun$ 
evolve through the instability strip in the H-R diagram \citep{te06}. 
Parabolic trends in Cepheid O--C diagrams --- temporal plots of the 
differences between Observed and Computed times of light maxima --- are 
diagnostic features of stars undergoing slow changes in mean radius 
\citep{pa58,st59}.

The case for the 3.7-day Cepheid RT~Aur is most unusual. Summaries by 
\citet{sz77,sz91} and \citet{fe93} of observed times of maximum light 
between 1897 and 1980 provide a strong case for a regular period decrease 
in the Cepheid \citep{tu98}, although \citet{sz77} preferred to interpret 
the O--C data as evidence for a discontinuous period change, contrary to 
the arguments for evolution \citep{sz83,te06}. The available O--C data to 
1980, from \citet{sz77,sz91}, \citet{fe93}, \citet{wu92}, and an 
unpublished list by Vitaly Goransky of the Sternberg Astronomical 
Institute, cited by \citet{ko06}, of observed times of maximum light 
are shown in Figure 1 (upper). The weighting scheme for the data used 
throughout this paper is that employed by \citet{sz77}, with weights 
assigned to sources not cited by \citet{sz77,sz91} on the basis of the 
perceived quality of the result. The negative parabolic trend is the 
signature of a regular period decrease \citep{st59}, and the inferred 
rate of $-0.123 \pm 0.018$ s yr$^{-1}$ is close to what is predicted 
from stellar evolutionary models for a star in the second crossing of 
the Cepheid instability strip \citep{te06}.

Regular photoelectric monitoring of the brightness variations of RT~Aur 
by professional observers ceased over a decade ago, with the exception 
of \citet{be97}, \citet{ba97}, \citet{ki98}, and observations by the 
Hipparcos satellite \citep{esa97}. A variety of other observations of 
the star, primarily by amateur astronomers, has generated additional 
times of light maximum that are listed by \citet{wu92}, \citet{ko06}, and 
\citet{me04,me06}. They are plotted in Figure 1 (lower), using the 
weighting scheme described above, but are generally of lower quality 
than most observations by professional observers, as indicated by the 
larger scatter in the more recent times of light maxima. Nevertheless, 
it is clear that current O--C data do not confirm the regular period 
decrease evident prior to 1980. A similar trend is indicated in an O--C 
plot for RT~Aur generated by \citet{be03} using low quality observations 
by members of the American Association of Variable Star Observers 
(AAVSO), who, in conjunction with European observers, have been the 
primary observers of the Cepheid in the current era.

Here we present additional data that confirm the more recent observations 
of the curious change in pulsation period for RT~Aur, and argue that the 
long-term brightness changes of the Cepheid are actually more consistent 
with a period {\it increase} than a period decrease. There is, in fact, 
convincing evidence for a superposed sinusoidal trend that hints at more 
complex behavior generally consistent with orbital motion in a binary 
system. The main point to be made, however, is that the true situation will 
only be established by further monitoring of the star. The last few decades 
of observation merely hint at the interesting changes occurring in the 
system.

\section{Observational Data}

We obtained a selection of new O--C data points for RT~Aur through analysis 
of a variety of unpublished observations for the Cepheid, which include 
observations by members of the American Association of Variable Star 
Observers (AAVSO) from \citet{he06} and data from Group ``Betelgeuse.'' 
The latter include individual observations of RT~Aur by group members, as 
well as data obtained from visual inspection of photographic images in the 
plate archives at Minsk and Odessa.

All data listed only by Julian Date were converted into heliocentric 
equivalents, and were phased using a new ephemeris given by:
\begin{equation}
{\rm HJD}_{\rm max} = 2441723.6925 + 3.72824 \: E ,
\end{equation}
where $E$ is the number of elapsed cycles. We also made use of a standard 
light curve for RT~Aur, in $B$ and $V$, constructed from the detailed 
photometry of \citet{wi73} supplemented by data from \citet{mb84} that 
were matched in both phase and magnitude to the observations of \citet{wi73}.

Since RT~Aur is a fifth magnitude Cepheid, its brightness is typically 
monitored optically by means of binoculars or low power oculars, although 
magnitude estimates without optical aid would likely be more accurate 
given the eye's increasing precision for estimating brightness levels when 
functioning near the visibility limit \citep{tu00}. Stellar brightness is 
more difficult to establish optically when it falls well above the eye 
limit, which may partly explain the large scatter in the AAVSO estimates 
for RT~Aur \citep{he06}, as indicated in Figure 2. The large number of 
individual AAVSO estimates compensates for the large scatter, however, and 
results in very precise O--C estimates. The AAVSO database for RT~Aur is 
relatively sparse prior to 1969, however \citep[cf.,][]{be03}, which 
restricts its usefulness mainly to the last four decades.

The Group ``Betelgeuse'' brightness estimates for RT~Aur are a mix of 
different sources: individual eye estimates obtained between 1989 and 2000, 
as well as from 2005 to 2007, by individual group members, using low power 
oculars and wide field telescopes, and eye estimates from photographs in 
the plate archives of Odessa and Minsk. The archival photographic material 
dates from 1988 to 1996, and consists of panchromatic GZS-2 film with a 
magnitude limit of $V = 9.0$, and A500 film exposed through a UV filter 
with a magnitude limit of $B = 9.0-9.5$. Typically 2--4 reference stars 
differing in brightness by $\sim 0.5$ magnitude were used for comparison 
purposes, with individual estimates made using the step method. Some 
typical light curves are illustrated in Figure 3, where the superior quality 
of eye estimates from single observers over those of inhomogeneous groups is 
evident.

We also obtained new $V-$band photometry for RT~Aur during January, 
February, and March 2007 using a ST9 CCD camera equipped with Bessel 
filters on the 0.28-m C11 Schmidt-Cassegrain telescope of the automated 
Abbey Ridge Observatory of Dave Lane. The data were normalized using the 
previously-constructed standard light curve. The observations are listed 
in Table 1, along with phases computed as indicated previously.

For reference purposes, we list in Table 2 the times for light maximum 
compiled by Goransky and not compiled elsewhere in the literature. Some 
of the cited values are of indeterminate authorship.

\section{Analysis}

Seasonal light curves for RT~Aur were constructed from the observational 
data, and were matched to the standard $B$ and $V$ light curves using 
the robust software described previously \citep{tu98}. Despite the large 
amount of scatter in the AAVSO observations, the large number of individual 
estimates results in relatively precise O--C estimates, as indicated by 
the small scatter for the AAVSO values in Figure 4 (upper). The individual 
light curves from the Group ``Betelgeuse'' data exhibit slightly smaller 
scatter, but generally result in less accurate O--C values because of the 
smaller number of individual estimates, according to Figure 4 (lower). Both 
sets of observations confirm the trend indicated by the O--C data derived 
by other observers, primarily amateur astronomers (Figure 1, lower). It is 
not clear from the O--C estimates cited by \citet{wu92}, \citet{ko06}, and 
\citet{me04,me06} how the times of light maximum were derived, but the 
techniques are apparently less robust than the variant of Hertzsprung's 
method employed here.

The new photometry obtained here (Table 1), as well as visual observations 
of the Cepheid by Bryukhanov between December 2006 and April 2007, also 
display a phase shift relative to the standard light curve, as evident from 
Figure 5, that confirms the O--C trend of the other observations. A 
compilation of all O--C estimates from the present study is given in Table 
3, including, where possible, reworkings of older data sets available in 
the literature.

The complete set of O--C data, including the values compiled by 
\citet{sz77,sz91}, that of \citet{ke71} cited by \citet{fe93}, \citet{wu92}, 
\citet{ko06} as given in Table 2, and \citet{me04,me06} is illustrated in 
Figure 6 (lower), relative to the situation that existed prior to 1980 
(Figure 6, upper). A weighted least squares fit of a parabola to the full 
data set indicates that RT~Aur is undergoing an overall period {\it increase} 
rather than a period decrease, at a calculated rate of $+0.082 \pm 0.012$ s 
yr$^{-1}$. The value is consistent with the 3.7-day pulsation period of 
RT~Aur, as indicated by its location in the period change diagram of 
Figure 7, which is adapted from Fig. 5 of \citet{te06}. RT~Aur has a 
pulsational amplitude near the maximum value displayed by Cepheids with 
periods of $\sim4$ days, so must lie near the center of the instability 
strip, in fact slightly towards the hot edge from strip center \citep{te06}. 
The location of the O--C datum for RT~Aur in Figure 7 is almost exactly 
that expected for a 3.7-day Cepheid in the third crossing of the instability 
strip lying slightly blueward of strip center.

It is possible to remove the parabolic evolutionary trend in the O--C data 
of Figure 6 (lower), and also correct for errors in the adopted ephemeris. 
The resulting O--C residuals for RT~Aur are plotted in Figure 8, and are 
analyzed below.

The sinusoidal trend of the O--C data residuals for RT~Aur is a feature 
observed in a few other Cepheids. In some cases such trends arise from 
random fluctuations in pulsation period for the stars, e.g. SV~Vul 
\citep{tb01,tb04}. One can test for the effect by analyzing the residuals 
using the procedure developed by \citet{ed29}, see \citet{tb01}. One 
examines the temporal differences $a(r)$ of each $r^{\rm th}$ observed 
light maximum residual from the null relation to compute the accumulated 
delays $u(x) = a(r+x) - a(r)$ between maxima separated by $x$ cycles. 
According to \citet{ed29}, the average value $\langle\;u(x)\;\rangle$ for 
the accumulated delays between light maxima separated by $x$ cycles, without 
regard for sign, is correlated with any random fluctuations in period $e$ 
by:
\begin{displaymath}
\langle\;u(x)\;\rangle^2 = 2 a^2 + x e^2 ,
\end{displaymath}
where $a$ is the size of the random errors in the measured times of light 
maximum.

For RT~Aur the results over 1000 cycles (not shown) yield a best-fitting 
weighted relation given by:
\begin{displaymath}
\langle\;u(x)\;\rangle^2 = 0.017 (\pm 0.023) + 0.0000 (\pm 0.0001) x .
\end{displaymath}
The zero-point for the relation, $a = 0.092\pm0.108$, implies uncertainties 
in the calculated times of light maximum of order $\pm0.34$ day ($\sim8$ 
hours), which is reasonable although significantly larger than the 
uncertainties generated by Hertzsprung's method. The slope of the relation 
corresponds to a value for the randomness parameter of magnitude $e = 0.002 
\pm0.006$, consistent with a null result. It appears that the sinusoidal 
trend in the O--C residuals for RT~Aur cannot be attributed to random 
fluctuations in period, according to an Eddington test performed on the 
observational data.

Alternatively, the trend may arise from light travel time effects in a 
binary system. The O--C residuals were examined for periodicity through a 
Fourier analysis, which produced a strong, well-defined signal for 
$P = 26,429\pm89$ days, or $72.36\pm0.24$ years. The data phased to that 
period and an arbitrary zero-point of HJD 2410000 are shown in Figure 9 
(upper). A least squares fit of a sine wave to the data gives a value of 
$a_1\;{\rm sin}\;i = 0.0619\pm0.0090$ light day $= 10.72\pm1.56$ A.U. 
$ = 1.60\;(\pm0.23)\times10^9$ km for the orbit of the Cepheid 
about the system barycenter. Of course, the orbit need not be circular; 
the adoption of $e = 0$ in the analysis was predicated by the scatter in 
the O--C residuals and the lack of solid evidence for a non-sinusoidal trend.

The sine wave solution also yields a mass function for the putative binary 
system of $M_2^3\;{\rm sin}\;i\;(M_1+M_2)^{-2} = 0.236\pm0.059\;M_{\sun}$. 
Such a large mass function implies a relatively high mass for the companion, 
as well as a strong likelihood that the orbit is nearly edge-on. With a 
mass of $M_1 = 4.7\pm0.3\;M_{\sun}$ for a fundamental mode Cepheid with 
the pulsation period of RT~Aur \citep{tu96}, the implied minimum mass for 
the secondary is of order $M_2 = 2.25\pm0.35\;M_{\sun}$, typical of a B9-A0 
dwarf. Such a large mass for the companion is ruled out, however, both by 
the color variations of the Cepheid \citep{le86}, which display no indication 
of a blue secondary, and by its ultraviolet spectrum \citep{ev92}, the latter 
indicating that any main sequence secondary for RT~Aur must be cooler than 
spectral type A4, or $\sim1.7\;M_{\sun}$. Conceivably there is an 
additional factor affecting the O--C variations other than random 
fluctuations in period or light travel time effects.

Radial velocity observations may provide a resolution to the paradox. 
\citet{sz91} has summarized the available systemic velocities for RT~Aur 
to 1991, to which we have added additional measures from the radial 
velocities tabulated by \citet{go98} and \citet{ki00}, with pulsational 
variations removed. The combined data phased to the ephemeris adopted 
for the O--C residuals are plotted in Figure 9 (lower). For orbital 
motion the radial velocity variations are a quarter cycle out of step 
with the O--C residuals, and leading them, so a sine wave with those 
characteristics was crudely fit by eye to the observations. The expected 
radial velocity half-amplitude according to the orbital solution is 
$\sim4.4$ km s$^{-1}$, but the observations appear to permit only a 
smaller value that we estimate as $K = 2.5\pm0.5$ km s$^{-1}$, with an 
implied systemic velocity of $18.4$ km s$^{-1}$. The projected orbital 
radius for the primary in this case is $a_1\;{\rm sin}\;i = 
9.09\;(\pm1.82)\times10^8$ km $= 6.07\pm1.21$ A.U., which results in 
a mass function of $M_2^3\;{\rm sin}\;i\;(M_1+M_2)^{-2} = 
0.043\pm0.015\;M_{\sun}$.

The radial velocity solution implies a minimum mass for the secondary of 
order $M_2 = 1.15\pm0.25\;M_{\sun}$, typical of a F7 dwarf. Such a solution 
is permitted by the lack of a companion detected through color variations 
and ultraviolet spectra, but rests upon an incomplete radial velocity 
solution. By chance the archival radial velocity observations of RT Aur 
are roughly coincident in orbital phase with more modern measurements, so 
only a third of the orbital cycle is covered observationally. There is 
also a potential zero-point offset for the earliest observations, which 
are those of \citet{du08}, remeasured by \citet{pe34} with similar 
results. The 1908 measures agree with the trend of the other radial 
velocity data only if they are systematically $\sim3$ km s$^{-1}$ too 
positive. Given that zero-point offsets of order $1-2$ km s$^{-1}$ are 
present even in some modern radial velocity measurements, a correction of 
that amount seems reasonable. Of course, it is conceivable that the 
effect may also indicate the presence of a third star in the system, 
but that is difficult to test with the available data. Certainly an 
improved spectroscopic orbital solution is only possible with a focused 
observational spectroscopic program on RT~Aur over the next half 
century, clearly a challenging task.

\section{Discussion}

The value of continued monitoring of Cepheid variables in an era when 
professional observations of such stars are declining is illustrated 
clearly by the case of RT~Aur. Circa 1993 when Fernie reviewed the 
situation \citep{fe93}, the available observations implied a regular 
period decrease for the Cepheid. Yet observations since then imply 
exactly the opposite: RT~Aur appears to be undergoing a regular period 
increase. The calculated rate of $+0.082 \pm 0.012$ s yr$^{-1}$is 
exactly that expected for a Cepheid in the third crossing of the 
instability strip lying near strip center, despite a superposed 
sinusoidal trend in the O--C data implying an additional complication.

The possibility that RT~Aur is undergoing random fluctuations in pulsation 
period is eliminated by an Eddington test on the residuals. The trend is 
consistent, however, with light time effects expected if RT~Aur is orbiting 
an unseen companion. The inferred minimum mass for the unseen companion is 
of order $2.25\pm0.35\;M_{\sun}$ from the O--C residuals, but only of order 
$1.15\pm0.25\;M_{\sun}$ according to the orbital radial velocity variations. 
The latter value is consistent with the lack of any evidence for a hot 
companion evident in the Cepheid's color variations and ultraviolet spectra. 
Additional observations of the star, and spectroscopic measurements in 
particular, may provide a more definitive estimate for the companion's 
characteristics.

The remarkable change in the O--C trend for RT~Aur, namely the switch from 
a period decrease prior to 1980 to a dominant period increase since then, 
is unusual but not without precedent. The 23-day Cepheid WZ Car, for 
example, appears to have changed from a regular period increase prior to 
1973 to a regular period decrease since then \citep{te03}, while the 4-day 
Cepheid Polaris underwent an astonishing glitch in its regular period 
increase circa 1963-66 \citep{te05} that is difficult to explain. Other 
surprises may be in store when a complete sample of Cepheid period changes 
is examined.

\acknowledgments

We acknowledge with thanks the variable star observations from the AAVSO 
International Database contributed by observers worldwide and used in this 
research. We are also grateful to the referee, Laszlo Szabados, for several 
useful suggestions that provided greater depth to the original study, and to 
Nicolai Samus and Vitaly Goransky for information on archival observations 
of RT Aur.

\clearpage
\begin{deluxetable}{ccc}
\tabletypesize{\scriptsize}
\tablecaption{CCD Observations of RT~Aurigae. \label{tbl-1}}
\tablewidth{0pt}
\tablehead{
\colhead{HJD} &\colhead{Phase} &\colhead{$V$}}

\startdata
2,454,114.4960 &0.504 &5.601 \\
2,454,122.5157 &0.655 &5.766 \\
2,454,122.6002 &0.678 &5.791 \\
2,454,123.4933 &0.917 &5.707 \\
2,454,124.5660 &0.205 &5.231 \\
2,454,124.8261 &0.275 &5.300 \\
2,454,128.5226 &0.266 &5.318 \\
2,454,128.6473 &0.300 &5.360 \\
2,454,135.5390 &0.148 &5.184 \\
2,454,135.6977 &0.191 &5.228 \\
2,454,136.6442 &0.445 &5.554 \\
2,454,167.5417 &0.732 &5.778 \\
2,454,183.6046 &0.041 &5.051 \\
\enddata

\end{deluxetable}

\clearpage
\begin{deluxetable}{rrrcl}
\tabletypesize{\scriptsize}
\tablecaption{Archival O--C Data for RT~Aurigae. \label{tbl-2}}
\tablewidth{0pt}
\tablehead{
\colhead{HJD$_{\rm max}$} &\colhead{Cycles} &\colhead{$O-C$} 
&\colhead{Weight} &\colhead{Source} \\
&\colhead{$E$} &\colhead{(days)} & & }

\startdata
2,417,173.360 &--6585 &+0.128 &0.5 &\citet{wi05} \\
2,418,347.279 &--6270 &--0.348 &0.0 &\citet{me41} \\
2,420,957.478 &--5570 &+0.082 &1.0 &\citet{ku35} \\
2,422,784.241 &--5080 &+0.008 &1.0 &\citet{so22} \\
2,436,146.219 &--1496 &--0.027 &1.0 &\citet{st71} \\
2,442,838.410 &+299 &--0.026 &0.5 &\citet{bo82} \\
2,443,181.270 &+391 &--0.164 &0.5 &\citet{bu81} \\
2,443,490.870 &+474 &--0.008 &0.5 &\citet{bu81} \\
2,443,535.540 &+486 &--0.077 &0.5 &\citet{bu81} \\
2,443,550.370 &+490 &--0.160 &0.5 &\citet{bu81} \\
2,443,990.420 &+608 &--0.042 &1.0 &\citet{ha80} \\
2,446,488.580 &+1278 &+0.197 &0.5 &\citet{go88} \\
2,446,518.405 &+1286 &+0.196 &0.5 &\citet{go88} \\
2,446,824.299 &+1368 &+0.374 &0.0 &\citet{go88} \\
2,446,827.900 &+1369 &+0.247 &0.5 &\citet{go88} \\
2,446,850.298 &+1375 &+0.276 &0.5 &\citet{go88} \\
2,446,917.304 &+1393 &+0.173 &1.0 &\citet{an87} \\
2,447,148.220 &+1455 &--0.062 &0.5 &\citet{rs88} \\
2,447,152.230 &+1456 &+0.220 &1.0 &\citet{an87} \\
2,447,234.249 &+1478 &+0.218 &1.0 &\citet{an87} \\
\enddata

\end{deluxetable}

\clearpage
\begin{deluxetable}{rrrccl}
\tabletypesize{\scriptsize}
\tablecaption{New O--C Data for RT~Aurigae. \label{tbl-3}}
\tablewidth{0pt}
\tablehead{
\colhead{HJD$_{\rm max}$} &\colhead{Cycles} &\colhead{$O-C$} 
&\colhead{Weight} &\colhead{Data Pts} &\colhead{Source} \\
&\colhead{$E$} &\colhead{(days)} & &\colhead{$n$} & }

\startdata
2,421,300.634 &--5478 &+0.240 &0.5 &43 &AAVSO \citep{he06} \\
2,421,647.171 &--5385 &+0.051 &0.5 &29 &AAVSO \citep{he06} \\
2,422,747.108 &--5090 &+0.157 &0.5 &25 &AAVSO \citep{he06} \\
2,426,419.438 &--4105 &+0.170 &2.0 &18 &\citet{du47} \\
2,426,971.249 &--3957 &+0.202 &2.0 &46 &\citet{du47} \\
2,427,504.400 &--3814 &+0.215 &1.0 &6 &\citet{du47} \\
2,429,249.163 &--3346 &+0.161 &3.0 &204 &\citet{be41} \\
2,429,625.692 &--3245 &+0.138 &3.0 &98 &\citet{be41} \\
2,432,955.026 &--2352 &+0.154 &3.0 &14 &\citet{eg57} \\
2,434,405.262 &--1963 &+0.104 &3.0 &9 &\citet{eg57} \\
2,435,821.997 &--1583 &+0.108 &3.0 &40 &\citet{pr61} \\
2,437,126.806 &--1233 &+0.034 &3.0 &10 &\citet{mi64} \\
2,438,010.307 &--996 &--0.059 &2.0 &4 &\citet{wi66} \\
2,438,424.195 &--885 &--0.005 &3.0 &13 &\citet{wj68} \\
2,439,132.508 &--695 &--0.058 &3.0 &20 &\citet{ta69} \\
2,439,147.429 &--691 &--0.050 &3.0 &19 &\citet{wj68} \\
2,440,079.464 &--441 &--0.075 &0.5 &16 &AAVSO \citep{he06} \\
2,440,351.595 &--368 &--0.105 &0.5 &50 &AAVSO \citep{he06} \\
2,440,675.924 &--281 &--0.133 &0.5 &57 &AAVSO \citep{he06} \\
2,440,981.756 &--199 &--0.016 &2.0 &20 &\citet{fm80} \\
2,440,996.641 &--195 &--0.045 &3.0 &5 &\citet{ev76} \\
2,441,030.238 &--186 &--0.001 &0.5 &44 &AAVSO \citep{he06} \\
2,441,250.206 &--127 &+0.000 &3.0 &88 &\citet{wi73} \\
2,441,705.070 &--5 &+0.019 &3.0 &20 &\citet{sz77} \\
2,441,854.201 &+35 &+0.020 &1.0 &177 &AAVSO \citep{he06} \\
2,442,137.473 &+111 &--0.054 &0.5 &119 &AAVSO \citep{he06} \\
2,442,525.277 &+215 &+0.013 &0.5 &234 &AAVSO \citep{he06} \\
2,442,920.412 &+321 &--0.045 &0.5 &179 &AAVSO \citep{he06} \\
2,443,241.186 &+407 &+0.100 &0.5 &123 &AAVSO \citep{he06} \\
2,443,539.286 &+487 &--0.059 &2.0 &7 &\citet{mb84} \\
2,443,975.518 &+604 &--0.031 &3.0 &23 &\citet{mb84} \\
2,444,135.795 &+647 &--0.069 &0.5 &99 &AAVSO \citep{he06} \\
2,444,378.093 &+712 &--0.106 &0.5 &171 &AAVSO \citep{he06} \\
2,444,534.792 &+754 &+0.007 &3.0 &7 &\citet{eg85} \\
2,444,758.522 &+814 &+0.042 &1.0 &147 &AAVSO \citep{he06} \\
2,445,108.977 &+908 &+0.043 &1.0 &190 &AAVSO \citep{he06} \\
2,445,463.189 &+1003 &+0.072 &1.0 &168 &AAVSO \citep{he06} \\
2,445,835.952 &+1103 &+0.011 &1.0 &153 &AAVSO \citep{he06} \\
2,446,190.169 &+1198 &+0.045 &1.0 &159 &AAVSO \citep{he06} \\
2,446,398.941 &+1254 &+0.035 &2.0 &23 & Minsk archives \\
2,446,563.027 &+1298 &+0.079 &1.0 &166 &AAVSO \citep{he06} \\
2,446,842.574 &+1373 &+0.008 &2.0 &26 &Bryukhanov \\
2,446,935.870 &+1398 &+0.098 &1.0 &161 &AAVSO \citep{he06} \\
2,447,308.614 &+1498 &+0.018 &1.0 &182 &AAVSO \citep{he06} \\
2,447,465.342 &+1540 &+0.160 &3.0 &27 &\citet{ba97} \\
2,447,562.026 &+1566 &--0.091 &1.0 &54 &Odessa archives \\
2,447,707.593 &+1605 &+0.076 &1.0 &124 &AAVSO \citep{he06} \\
2,447,942.456 &+1668 &+0.059 &1.0 &92 &Sergey \\
2,447,949.853 &+1670 &+0.000 &1.0 &54 &Odessa archives \\
2,447,949.888 &+1670 &+0.035 &1.0 &314 &Kosa-Kiss et al. \\
2,448,061.730 &+1700 &+0.029 &1.0 &161 &AAVSO \citep{he06} \\
2,448,251.930 &+1751 &+0.089 &1.0 &69 &Sergey \\
2,448,270.569 &+1756 &+0.087 &1.0 &114 &Schukin, Sergey, Kosa-Kiss, Mamedov \\
2,448,304.135 &+1765 &+0.098 &2.0 &94 &Minsk archives \\
2,448,382.431 &+1786 &+0.102 &1.0 &131 &AAVSO \citep{he06} \\
2,448,490.52 &+1815 &+0.078 &3.0 &69 &Hipparcos \\
2,448,617.050 &+1849 &--0.158 &0.5 &28 &Sergey \\
2,448,632.137 &+1853 &+0.015 &1.0 &57 &Narkevich \\
2,448,662.024 &+1861 &+0.076 &1.0 &105 &Minsk archives \\
2,448,710.490 &+1874 &+0.076 &1.0 &44 &Grigorenko \\ 
2,448,796.305 &+1897 &+0.142 &1.0 &117 &AAVSO \citep{he06} \\
2,448,997.475 &+1951 &--0.014 &1.0 &53 &Narkevich \\
2,448,997.622 &+1951 &+0.133 &1.0 &53 &Minsk archives \\
2,449,016.242 &+1956 &+0.112 &1.0 &36 &Sergey \\
2,449,083.323 &+1974 &+0.084 &1.0 &289 &AAVSO \citep{he06} \\
2,449,269.725 &+2024 &+0.075 &1.0 &52 &Minsk archives \\
2,449,269.785 &+2024 &+0.135 &1.0 &52 &Kosinski \\
2,449,362.956 &+2049 &+0.100 &1.0 &48 &Sergey \\
2,449,530.673 &+2094 &+0.045 &0.5 &334 &AAVSO \citep{he06} \\
2,449,735.831 &+2149 &+0.151 &1.0 &68 &Minsk archives \\
2,449,888.560 &+2190 &+0.022 &0.5 &362 &AAVSO \citep{he06} \\
2,450,015.384 &+2224 &+0.086 &3.0 &9 &\citet{be97} \\
2,450,112.202 &+2250 &--0.031 &1.0 &55 &Minsk archives \\
2,450,164.357 &+2264 &--0.071 &1.0 &64 &Sergey \\
2,450,198.098 &+2273 &+0.116 &3.0 &19 &\citet{ki98} \\
2,450,242.841 &+2285 &+0.120 &1.0 &418 &AAVSO \citep{he06} \\
2,450,511.291 &+2357 &+0.137 &1.0 &343 &AAVSO \citep{he06} \\
2,450,757.365 &+2423 &+0.147 &1.0 &156 &AAVSO \citep{he06} \\
2,450,865.339 &+2452 &+0.002 &1.0 &47 &Sergey \\
2,450,884.076 &+2457 &+0.098 &1.0 &283 &AAVSO \citep{he06} \\
2,451,126.393 &+2522 &+0.079 &1.0 &206 &AAVSO \citep{he06} \\
2,451,245.622 &+2554 &+0.004 &0.5 &597 &AAVSO \citep{he06} \\
2,451,488.012 &+2619 &+0.059 &1.0 &110 &AAVSO \citep{he06} \\
2,451,674.434 &+2669 &+0.069 &1.0 &333 &AAVSO \citep{he06} \\
2,452,036.067 &+2766 &+0.063 &1.0 &392 &AAVSO \citep{he06} \\
2,452,371.704 &+2856 &+0.158 &1.0 &269 &AAVSO \citep{he06} \\
2,452,815.341 &+2975 &+0.135 &1.0 &205 &AAVSO \citep{he06} \\
2,453,128.586 &+3059 &+0.207 &0.5 &232 &AAVSO \citep{he06} \\
2,453,438.082 &+3142 &+0.259 &1.0 &28 &Semenyuta \\
2,453,478.956 &+3153 &+0.122 &2.0 &57 &Balyuk \\
2,453,508.822 &+3161 &+0.163 &0.5 &282 &AAVSO \citep{he06} \\
2,453,780.985 &+3234 &+0.164 &0.5 &138 &AAVSO \citep{he06} \\
2,454,131.557 &+3328 &+0.282 &2.0 &13 &Abbey Ridge data \\
2,454,153.871 &+3334 &+0.226 &1.0 &45 &Bryukhanov \\
\enddata

\end{deluxetable}

\clearpage

\begin{figure}
\plotone{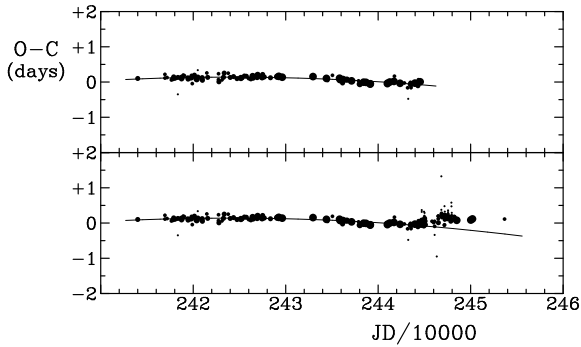}
\caption{O--C data for RT~Aur as available by 1980 (upper), and with more 
recently published observations (lower), with symbol size proportional to 
the weight assigned to each datum. The negative parabolic trend in both 
cases is a least squares fit to the pre-1980 data, indicative of a regular 
period decrease for the Cepheid. \label{fig1}}
\end{figure}

\begin{figure}
\plotone{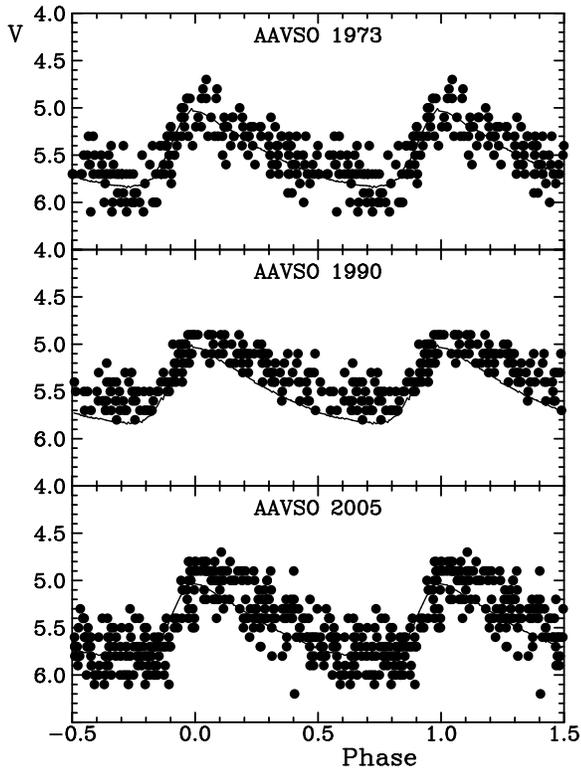}
\caption{A sample of yearly compilations of observations for RT~Aur taken 
from the AAVSO database. The plotted relation in each case is the adopted 
standard light curve. \label{fig2}}
\end{figure}

\begin{figure}
\plotone{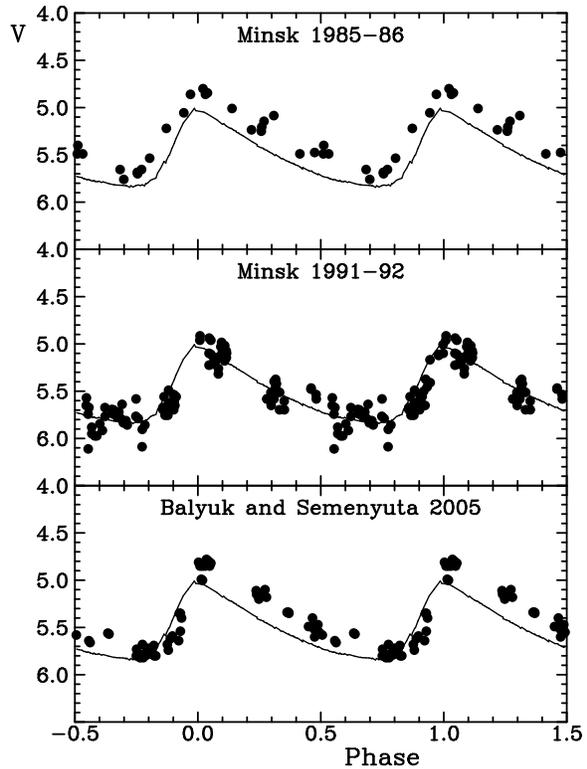}
\caption{A sample of observations for RT~Aur obtained from Group 
``Betelgeuse.'' The upper two sections contain data obtained from eye 
estimates off plates in the Minsk collection, the lower section individual 
observations by Group observers Balyuk and Semenyuta. The plotted relation 
in each case is the adopted standard light curve. \label{fig3}}
\end{figure}

\begin{figure}
\plotone{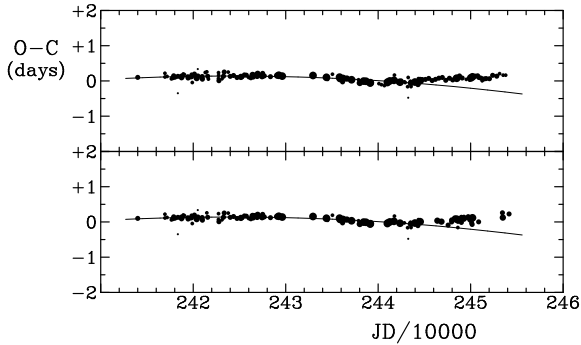}
\caption{Newly derived O--C data for RT~Aur plotted relative to existing 
pre-1980 data for the AAVSO sample (upper) and the Group ``Betelgeuse'' 
sample (lower). The plotted relation in each case is the least squares fit 
to the pre-1980 data, and symbol size is proportional to the weight assigned 
to the O--C datum. \label{fig4}}
\end{figure}

\begin{figure}
\plotone{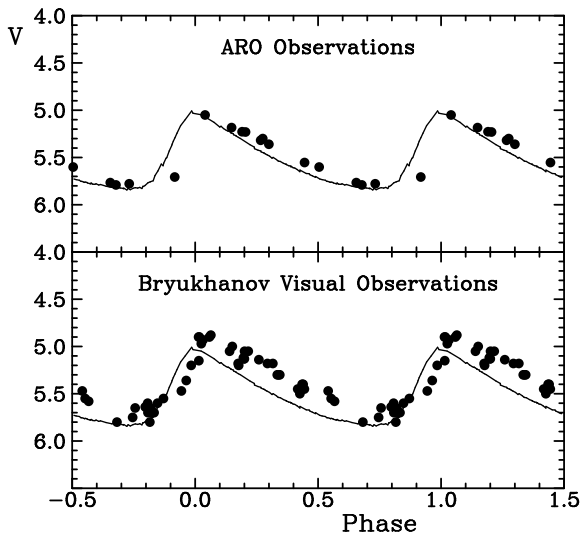}
\caption{New CCD observations of RT~Aur obtained from the Abbey Ridge 
Observatory (upper) and visual observations by Bryukhanov with a 
$7\times50$ monocular (lower). The plotted relations are the adopted 
standard light curve. \label{fig5}}
\end{figure}

\begin{figure}
\plotone{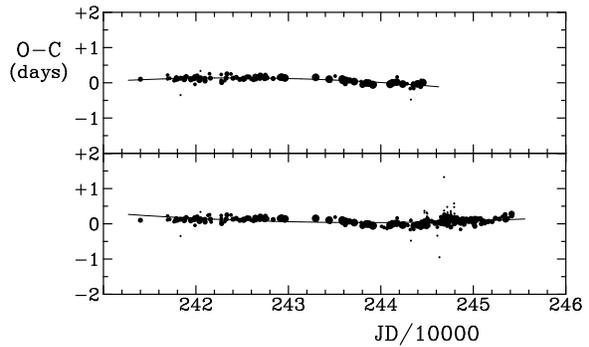}
\caption{The available O--C data for RT~Aur prior to 1980 (upper) and at 
present (lower), with symbol size as in Figs. 1 and 4. The negative 
parabolic trend (upper) and positive parabolic trend (lower) are least 
squares fits to the data in each case. The lower trend indicates a regular 
period {\it increase} for the Cepheid, with a superposed sinusoidal 
trend. \label{fig6}}
\end{figure}

\begin{figure}
\plotone{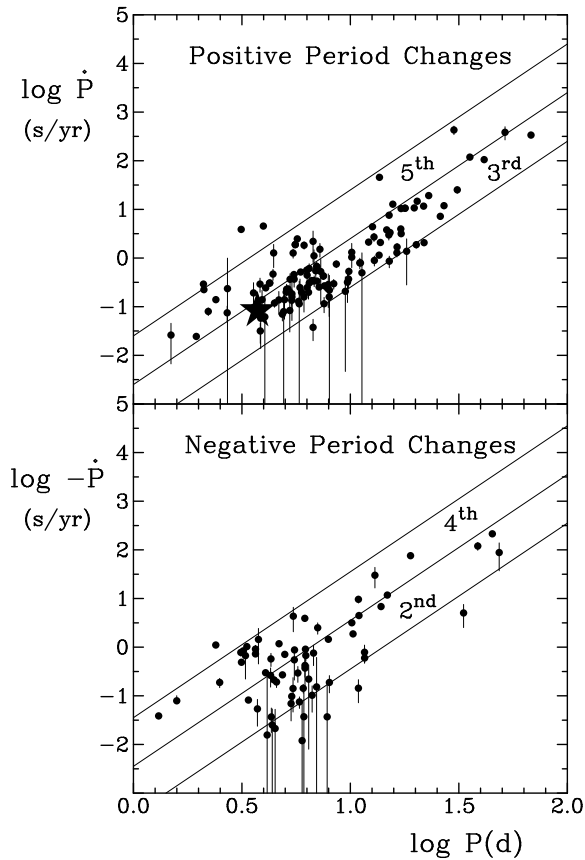}
\caption{Observed rates of period change for well-studied Galactic Cepheids 
\citep{te06}, with the present result for RT~Aur plotted as a star symbol. 
The plotted lines indicate the empirical delineation derived by \citet{te06} 
for different instability strip crossing modes. \label{fig7}}
\end{figure}

\begin{figure}
\plotone{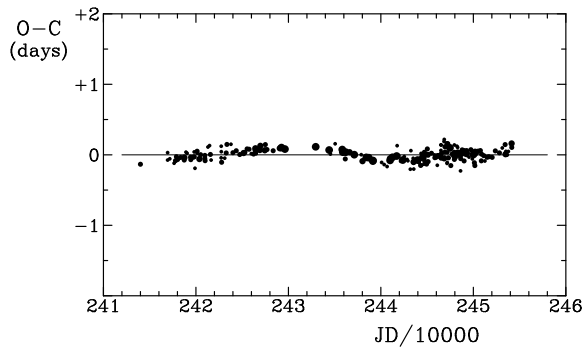}
\caption{The available O--C data residuals for RT~Aur with the parabolic 
evolutionary trend removed and corrected for errors in the adopted 
ephemeris. Symbol size is that used in Figs. 1, 4, and 6, except that 
zero-weight points are not plotted. \label{fig8}}
\end{figure}

\begin{figure}
\plotone{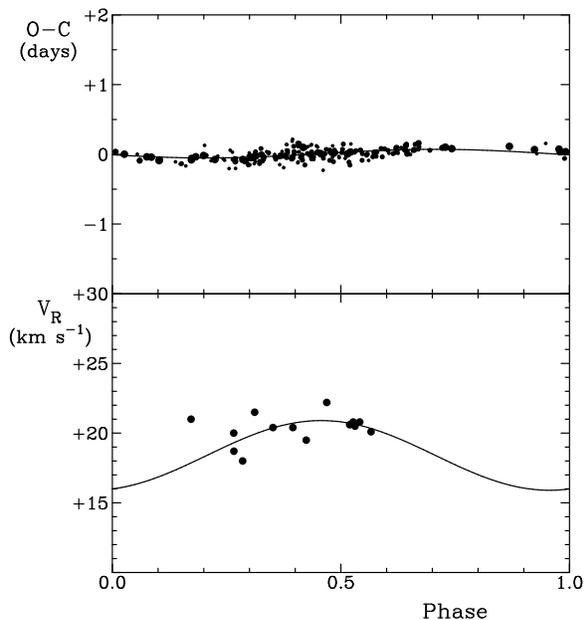}
\caption{Phased O--C data residuals (upper, with zero-weight points omitted) 
and systemic radial velocities (lower) for RT~Aur for an adopted zero-point 
epoch of HJD 2410000 and $P = 26,429$ days. Sine wave fits to the data are 
depicted, with an adopted quarter cycle offset in the lower plot. \label{fig9}}
\end{figure}

\end{document}